\documentclass[10pt,pra,aps,twocolumn,superscriptaddress,nofootinbib,floatfix,notitlepage]{revtex4-2}
\usepackage[latin1]{inputenc}
\usepackage{amsmath,color}
\usepackage[dvipsnames]{xcolor}
\usepackage{amsfonts}
\usepackage{amssymb}
\usepackage{bm}
\usepackage{appendix}
\usepackage{breqn}
\usepackage[usestackEOL]{stackengine}
\usepackage[percent]{overpic}
\usepackage{tikz}
\usepackage{braket}
\usepackage{hyperref}
\hypersetup{colorlinks=true,    linkcolor=blue,
        citecolor = red,  urlcolor=magenta}
\usepackage[english]{babel}

\begin{document}

\title{EIT spectra of Rydberg atoms dressed with dual-tone radio-frequency fields}\thanks{Publication of the U.S. government, not subject to U.S. copyright.}
\author{Maitreyi Jayaseelan}
\thanks{maitreyi.jayaseelan@colorado.edu}
\affiliation{Department of Physics\char`,\, University of Colorado\char`,\, Boulder\char`,\, CO 80302\char`,\, USA}
\author{Andrew P. Rotunno}
\author{Nikunjkumar Prajapati}
\author{Samuel Berweger}
\author{Alexandra B. Artusio-Glimpse}
\author{Matthew T. Simons}
\author{Christopher L. Holloway}
\thanks{christopher.holloway@nist.gov}
\affiliation{National Institute of Standards and Technology\char`,\, Boulder\char`,\, CO 80305\char`,\, USA}
\date{\today}
\begin{abstract} We examine spectral signatures of Rydberg atoms driven with near-resonant dual-tone radio-frequency (RF) fields in the regime of strong driving. 
We experimentally demonstrate and theoretically model a variety of nonlinear and multiphoton phenomena in the atomic Rydberg response that manifest in the EIT spectra. 
Our results echo previous studies of two-level atoms driven with bichromatic optical fields. 
In comparison to the optical studies, the RF-driven Rydberg system utilizes a more complex excitation pathway, and electromagnetic fields from two different spectral regimes: a two-photon optical excitation continuously creates highly excited Rydberg atoms, while RF fields drive resonant coupling between the Rydberg levels and generate strong mixing. 
Yet, our spectra reflect nearly identical effects of the dual-tone RF fields on the atomic Rydberg observables, showing detuning-dependent splittings and Rabi frequency dependent peak numbers and relative strengths, and avoided crossings at subharmonic resonances.
We thus validate previous two-state models in this more complex physical system.
In the context of Rydberg electrometry, we use these investigations to explore a technique where we tune a known RF field to observe spectra which give frequency and power of an unknown RF field using the complex dual-tone spectra.

\end{abstract}
	\maketitle

\section{Introduction} 

Two-level atoms driven by intense bichromatic optical fields have been  the subject of extensive experimental \cite{zhu_resonance_1990, yu_driving_1997, papademetriou_autler-townes_1996, PhysRevA.49.R1519} and theoretical \cite{freedhoff_resonance_1990, agarwal_spectrum_1991, PhysRevA.53.990, ficek_two-level_1996, van_leeuwen_autler-townes_1996, chien_quantum_1998, rudolph_multiphoton_1998}  investigation. In these systems, both resonance fluorescence and absorption spectra have shown physics beyond the single-frequency Rabi splitting characteristic of atoms subject to monochromatic driving, including detuning-dependent and Rabi frequency independent spectral splittings, subharmonic resonances, and phase-dependent atomic dynamics. These investigations point to a wealth of multiphoton dynamics that are accessible to systems driven by multiple frequencies: bichromatic electromagnetically induced transparency (EIT) has been demonstrated in both cold atoms \cite{wang_bichromatic_2003} and in hot vapors \cite{PhysRevA.87.055401}, and bichromatic and multifrequency fields have been employed in novel cooling methods for alkali atoms \cite{RevModPhys.89.041001}. Looking beyond atomic vapors, the spectra of bichromatically driven solid state systems with single-molecule impurities \cite{PhysRevLett.78.3673} and nuclear spins of nitrogen vacancy centers in diamond \cite{greentree_probing_1999} have revealed well-resolved subharmonic resonances and multiphoton effects, while a bichromatically driven quantum dot system 
demonstrated predicted quantum interference effects in fluorescence spectra \cite{ficek_quantum_1999, PhysRevLett.114.097402}.

In the radio-frequency (RF) and microwave domains (MHz and GHz), dual-tone dressing has been used to demonstrate a dynamically modulated Autler--Townes (AT) effect in superconducting qubits, providing an enhanced experimental toolbox for qubit manipulation and control \cite{pan_dynamically_2017}. It has also been investigated in the context of alignment-based magnetic resonance spectra in cesium, where dual-tone driving between the ground state hyperfine levels modifies the standard AT splitting of the system dressed by a single field \cite{PhysRevA.106.023108}. In Rydberg atoms, dual-tone microwave dressing was used to achieve a polarizability nulling effect  \cite{PhysRevA.97.012515}.
An aspect of Rydberg systems that has been less explored is their behaviour under near-resonant dual- and multi-tonal RF dressing. 

Over the last decade, continuously detected Rydberg EIT systems have proven to be an invaluable technology for sensitive, external calibration-free electrometry using the AT splitting (Rabi splitting) of Rydberg energy levels dressed by an RF field \cite{9748947}. Under single tone resonant driving, the AT spectra display splittings proportional to field strength, providing a direct measurement of electric field; the number and relative strengths of  spectral peaks are independent of field strength in this case. For strong driving with dual-tone RF dressing, multiphoton effects may be expected to yield spectra that are qualitatively different than those obtained with single tone driving.

Here we extend the physics of bichromatic  optical dressing of atoms to the RF regime, using a two-level Rydberg system probed by EIT in a warm atomic vapor of $^{85}$Rb. Atomic population is driven from the ground electronic state into a Rydberg state via an intermediate energy level using a two-photon optical excitation scheme. Two near-resonant RF fields couple this Rydberg state to an adjacent Rydberg state. While the full dynamics of this system is rather complex, we show that the experimental spectra may be modeled more simply by treating the two Rydberg levels as an isolated two-level system driven by a dual-tone electromagnetic field. We use a Floquet analysis  to model the response of the dressed Rydberg levels, obtaining good agreement with the rich experimental spectra. 

We emphasize that the experimental spectra are optically detected (EIT) Rydberg state energy spectra. The two-level dressed-atom physics of the RF-Rydberg system is probed with electromagnetic frequencies belonging to a different spectral range than those fields that create the multiphoton Floquet spectra. Our results thus validate the applicability of two-level dressed atom physics in the RF domain, where the optical fields that create the highly excited Rydberg atoms may in turn be regarded as indirect probes that do not significantly alter the relatively long-lived two-level system dynamics. Since the Rydberg system displays a wide range of resonances from GHz to MHz frequencies, these models may be validated for a wide range of dressing field frequencies. Further, the large electric dipole moments of highly excited Rydberg states allow the multiphoton dynamics of dual-tone dressed atoms to be demonstrated with modest RF field amplitudes when compared to optical dressing.

We distinguish our experiments from the ``atom mixer" configurations which used two RF tones applied to the atoms to transfer an intermediate frequency into the optical domain, providing phase sensitive detection for RF fields \cite{simons_rydberg_2019}. The atom mixer used a strong resonant local oscillator (LO) and a weaker signal field several kHz detuned and well within the EIT linewidth. Further, only the resulting beat signal on the optical frequency of the probe was detected. Here, we operate in the strong-driving regime, where both RF fields contribute non-trivially to the multiphoton dynamics that are integral to the spectra we observe.
We explore configurations where one or both fields are far ($\lesssim$\,200~MHz) off-resonance from the Rydberg transition. We further distinguish these experiments from previous bichromatic EIT configurations \cite{wang_bichromatic_2003, PhysRevA.87.055401} that used optical fields in a three-level $\Lambda$ configuration in contrast to our experiment where the dual-tone RF-induced Rabi splitting of two Rydberg levels is probed by a cascade EIT system, and from the dual-tone driving of atomic energy levels employed in Ref.~\cite{PhysRevA.106.023108} where the coupling is through the magnetic dipole term. 

\begin{figure}[t]
    \centering     \includegraphics[width=\linewidth,height=\textheight,keepaspectratio]{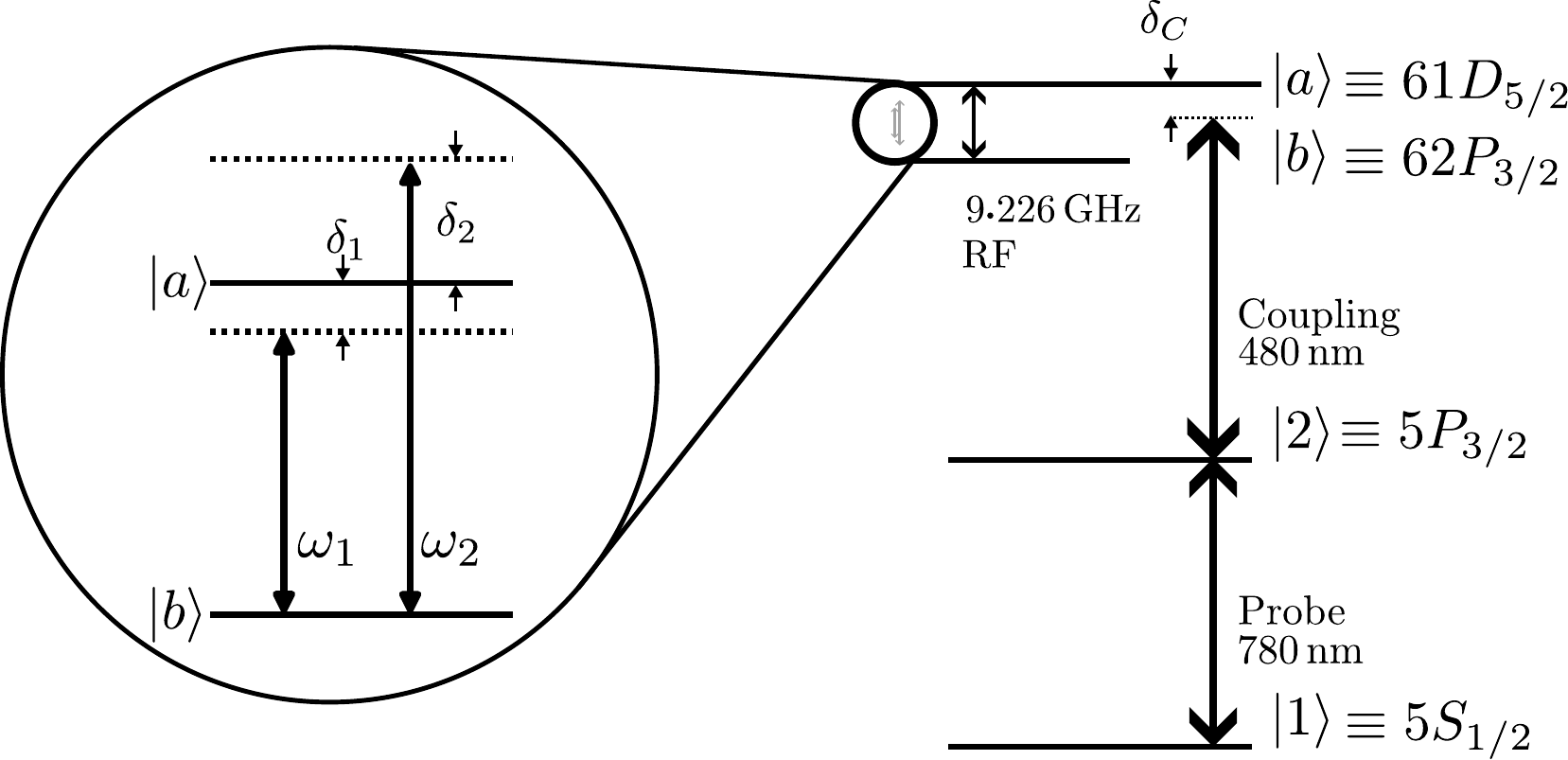}
    \caption{Rydberg atom with applied dual-tone RF fields.}
    \label{fig:1}
\end{figure}

We begin with a description of the experimental Rydberg EIT setup in Sec~\ref{sec:Experiment}, and present our dual-tone two-level Floquet theoretical model in Sec~\ref{sec:theory}. Experimental results are presented against computed spectra in Sec~\ref{sec:experimentalresults} for `symmetric cases,'  and for `asymmetric' cases in Sec~\ref{sec:applicationscenario}, with an eye toward applications. 
We conclude in Sec~\ref{sec:conclusion}.
Extensions of our model to include atomic fine structure and magnetic structure are in Appx.~\ref{sec:finestructure}. 

\section{Experiment}\label{sec:Experiment}
We use a two-photon excitation scheme to create excited Rydberg atoms in a $^{85}$Rb vapor cell at room temperature. The excitation pathway $5S_{1/2}\rightarrow 5P_{3/2} \rightarrow nD_{5/2}$  is shown in Fig.~\ref{fig:1}. We use a probe laser at $780$~nm, locked to the $F=3\rightarrow F'=4$ transition on the $D_2$ line of $^{85}$Rb, and a counter-propagating coupling laser at $480$~nm to excite atoms into the Rydberg state. 
We employ cascade EIT between the atomic ground state and the Rydberg state as our detection scheme: as the coupling laser is scanned through the Rydberg manifold, EIT of the probe beam appears when the two-photon system is resonant to a Rydberg state. 
Both lasers are power-locked with acousto-optic modulators. 
We isolate the EIT signal using differential detection of two power-balanced probe beams; one beam overlapped with the coupling laser for EIT, and the other an absorption reference. 

In this experiment, we investigate the response of the atomic Rydberg states to a dual-tone RF field addressing the Rydberg transition $61D_{5/2} \rightarrow 62P_{3/2}$ with a transition dipole moment $\wp= 2366\,e a_0$ and a resonant transition frequency of $9.226$~GHz, calculated using the ARC software package \cite{sibalic_arc_2017}. The Rydberg transition resonance frequency $\omega_0$ is verified using a single RF field at moderate power and balancing the two AT split peaks. RF fields are applied to the vapor cell using a horn antenna that is oriented such that that the RF fields propagate perpendicular to the direction of propagation of the optical fields. The RF fields and the probe and coupling lasers are all linearly polarized in the $\hat{z}$ direction, perpendicular to the plane of the optical table. The two RF tones whose effects we investigate in this work are outputs from a dual-output signal generator with independently controllable powers and detunings. The two outputs are combined with a power combiner and applied to the RF horn antenna.

For our data, we use a frequency scale set by a scan of the coupling laser detuning ($\delta_c$) over the states $61D_{3/2}$ and $61D_{5/2}$,  which have a fine-structure splitting of $50.339$~MHz, calculated using Ref.~\cite{sibalic_arc_2017}. These EIT scans, simultaneously collected in a reference cell away from the horn, also provide a frequency reference to correct for offsets of the scans due to laser drift and other environmental effects.
A slight residual `drift' towards negative $\delta_c$ is observed with increased field, potentially as a result of residual field-dependent shifting in the reference cell, which was poorly shielded from RF reflections. The two-level theory that we employ does not reproduce these shifts; however the spectral characteristics we emphasize in this work remain unaffected.

\section{Theoretical framework: Floquet Hamiltonian} \label{sec:theory}We now discuss a simplified theoretical model that we will use to analyse our experimental data. We restrict this discussion to the dynamics of a two-level atomic system composed of two Rydberg states dressed by a dual-tone RF field. The two-level model reproduces the main features of the experimental spectra with good agreement, which is one of the main results of this work. Extensions to the two-level model that consider atomic structure are discussed in Appx.~\ref{sec:finestructure}.

Consider two Rydberg states $|a\rangle$ and $|b\rangle$ dressed with a dual-tone RF field of the form \begin{align}
\mathbf{E}(t) = & \left(|E_1|\cos(\omega_1 t) + |E_2|\cos(\omega_2 t + \Phi) \right)\mathbf{\hat{z}}, 
\end{align} where $|E_i|$ is the magnitude of the field at frequency $\omega_i$, and $\Phi$ is the relative phase between the two fields. We denote the bare Rydberg atomic resonance as $\omega_0$, so that $\delta_1 = \omega_0 - \omega_1$ and $\delta_2 = \omega_0 - \omega_2$ are the detunings of the two field components from the Rydberg resonance. We note that the trigonometric identity $\cos(\theta)+\cos(\phi) = 2\cos\left(\frac{\theta+\phi}{2}\right)\cos\left(\frac{\theta-\phi}{2}\right)$ suggests the interpretation of our dual-tone setup as a carrier at the mean frequency, amplitude modulated at a rate of half their difference. 

Defining the Rabi frequencies of the two components as $\Omega_i = -\langle a|\mathbf{d}\cdot \mathbf{\hat{z}} | b\rangle |E_i|/\hbar $, where 
$\wp_{a,b} = \bra{a}\mathbf{d}\ket{b}$
is the atomic dipole moment and $\mathbf{\hat{z}}$ is the direction of linear polarization of the applied RF fields, the interaction Hamiltonian in the rotating wave approximation (RWA) is \cite{berman}:

\begin{equation}
    H_{\text{RWA}} = \frac{\hbar}{2}
    \begin{pmatrix}
        0 &&         
        \begin{matrix}
           \Omega_1 e^{-i\delta_{\delta} t/2}\nonumber \\
            + \Omega_2 e^{i(\delta_{\delta} t/2 + \Phi)}\nonumber
        \end{matrix} \\
        &&\\
        \begin{matrix}
           \Omega_1^* e^{i\delta_{\delta} t/2}\nonumber \\
           + \Omega_2^* e^{-i(\delta_{\delta} t/2 + \Phi)}\nonumber
        \end{matrix}  && -\Sigma_{\delta}
    \end{pmatrix},
\end{equation} where we define the difference and sum of the two RF detunings $\delta_{1,2}$:
\begin{align}
    \delta_{\delta} &\equiv \delta_2 - \delta_1 = \omega_2-\omega_1    \\
\Sigma_{\delta} &\equiv \delta_1 + \delta_2 = 2\omega_0 - (\omega_1+\omega_2)\,.
\end{align} 
We set $\Phi = 0$ in the subsequent analysis. Note that setting $|\Omega_1| = |\Omega_2|$ reduces $H_{\text{RWA}}$ to the lab-frame Hamiltonian of a two-level system with energy separation $\Sigma_{\delta}$ and coupled by a field at frequency $\delta_{\delta}/2$. 
A particular case is for  $\delta_1 = -\delta_2$, where the Floquet modulation frequency is equal to the detunings as $\delta_\delta/2=|\delta_1|=|\delta_2|$, and in the RWA the states are degenerate as $\Sigma_\delta=0$.

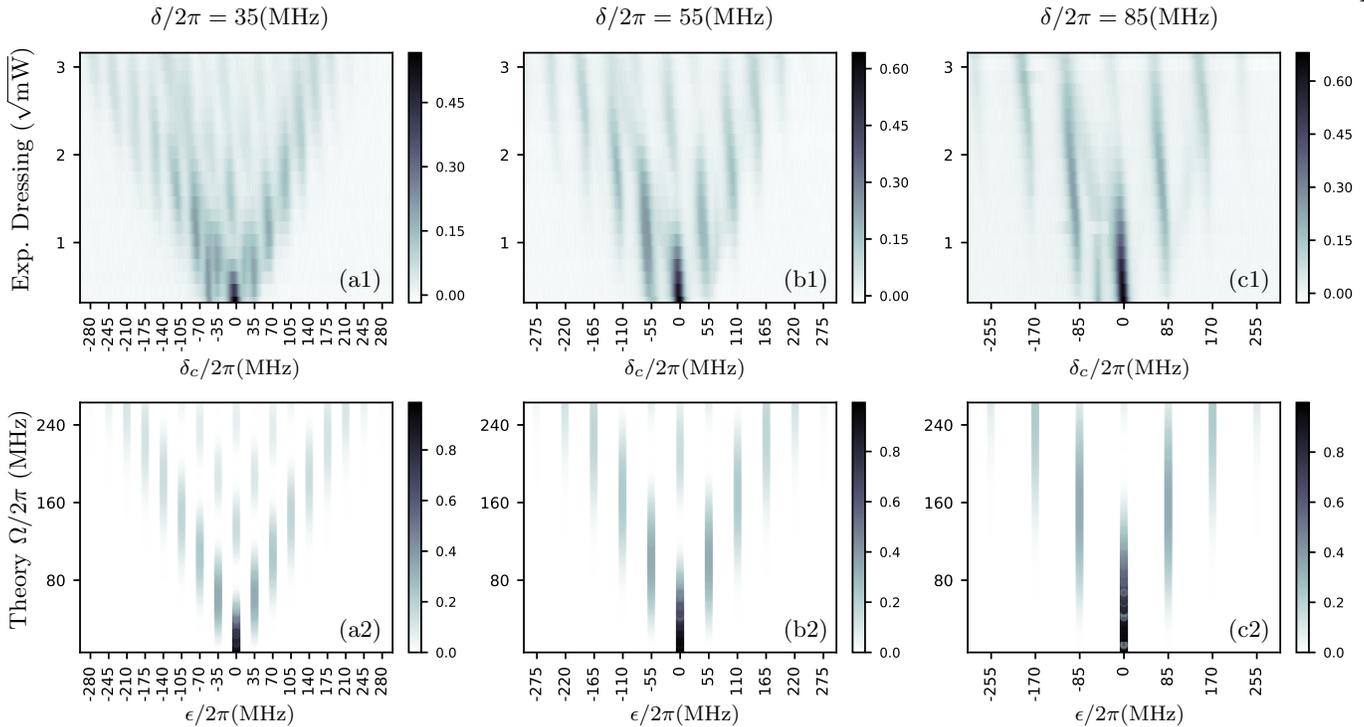
\begin{figure*}[htb]
    \centering     
    \begin{overpic}[width = \textwidth, tics=5, unit=1pt, grid = false, trim=0 0 0 0, clip]
    {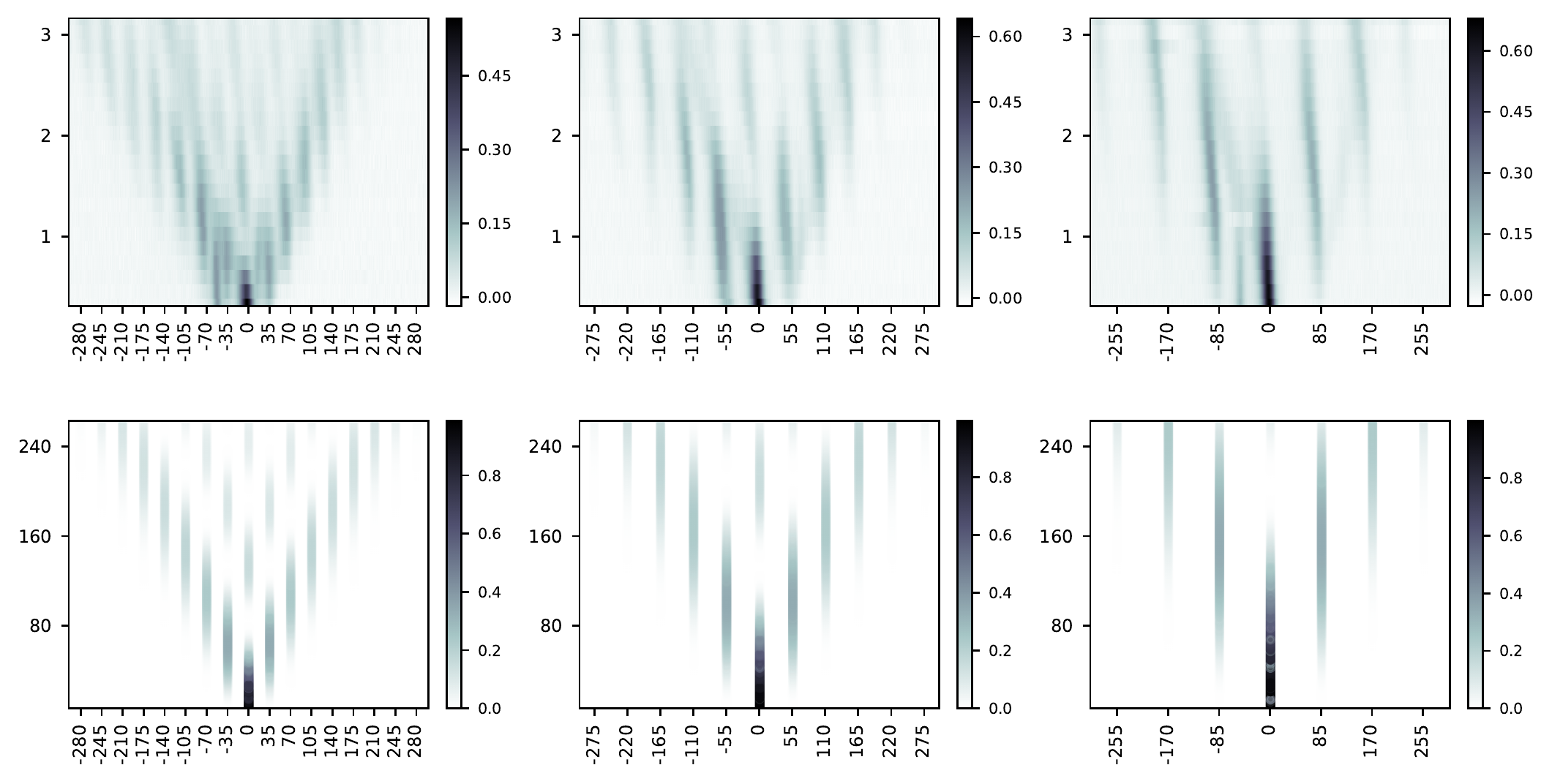}
    \put(23, 32){ \textcolor{black}{(a1)}   }
    \put(56, 32){ \textcolor{black}{(b1)}   }
    \put(89, 32){ \textcolor{black}{(c1)}   }
    \put(23, 6){ \textcolor{black}{(a2)}   }
    \put(56, 6){ \textcolor{black}{(b2)}   }
    \put(89, 6){ \textcolor{black}{(c2)}   }

    \put(16,25.5){\makebox[0pt]{ \footnotesize{\Centerstack{\textcolor{black}{$\delta_c/2\pi$(MHz)}   }}}}
    \put(49,25.5){\makebox[0pt]{ \footnotesize{\Centerstack{\textcolor{black}{$\delta_c/2\pi$(MHz)}   }}}}
    \put(82,25.5){\makebox[0pt]{ \footnotesize{\Centerstack{\textcolor{black}{$\delta_c/2\pi$(MHz)}   }}}}

    \put(16, -0.2){\makebox[0pt]{ \footnotesize{\Centerstack{\textcolor{black}{$\epsilon/2\pi$(MHz)}   }}}}
    \put(49, -0.2){\makebox[0pt]{ \footnotesize{\Centerstack{\textcolor{black}{$\epsilon/2\pi$(MHz)}   }}}}
    \put(82, -0.2){\makebox[0pt]{ \footnotesize{\Centerstack{\textcolor{black}{$\epsilon/2\pi$(MHz)}   }}}}

    \put(16, 51.5){\makebox[0pt]{ \Centerstack{\textcolor{black}{$\delta/2\pi = 35$(MHz)}   }}}
    \put(49, 51.5){\makebox[0pt]{ \Centerstack{\textcolor{black}{$\delta/2\pi = 55$(MHz)}   }}}
    \put(82, 51.5){\makebox[0pt]{ \Centerstack{\textcolor{black}{$\delta/2\pi = 85$(MHz)}   }}}

    \put(0, 40){\makebox[0pt]{\Centerstack{\rotatebox{90}{Exp. Dressing ($\sqrt{\text{mW}}$)}}}}
    \put(0, 14){\makebox[0pt]{\Centerstack{\rotatebox{90}{Theory $\Omega/2\pi$ (MHz)}}}}
        
    \end{overpic}
    \caption{Experimental (a1, b1, c1) and theoretical (a2, b2, c2) waterfall plots showing Rydberg EIT spectra obtained through a simultaneous scan of $|\Omega_1| = |\Omega_2| = |\Omega|$ with detunings $\delta_1 = -\delta_2 = \delta$ kept constant. The spectral features appear at coupling laser detunings spaced by the symmetric detuning $\delta$. The columns correspond to $\delta = 35$~MHz, $55$~MHz, and $85$~MHz. These locations are marked on the $x$ axis to emphasize the appearance of the spectral peaks at these locations. The features are reproduced in the theoretical Floquet quasienergy spectra, with waterfall plots over $\Omega$ showing the mode occupations of Floquet modes, against Floquet quasienergy $\epsilon$. The Floquet modes appear at quasienergies spaced by $\delta$ and this spacing remains constant as we scan $\Omega$. The mode occupation has a sensitive dependence on $\Omega$.}
    \label{figure:2}
\end{figure*}

Unlike the case of a two-level atom dressed by a monochromatic field, the dual-tone dressed system in the rotating frame shows a residual time dependence in the off-diagonal coupling terms, so that the Schrodinger equation for the system cannot be integrated directly. Nevertheless, the time-periodicity of the residual driving confers a symmetry that allows a conserved quasienergy for the system. We treat this residual time dependence using a Floquet picture, following Shirley's approach to obtaining the Floquet modes and quasienergies of the periodically modulated system \cite{shirley_solution_1965}. This approach promotes the time-dependent Hamiltonian from the state space $\mathcal{H}$ to an extended Floquet space where the eigenvectors are now labeled by two indices: the eigen index of the bare Hamiltonian and the ``photon number" index of the Floquet mode of $N$ photons with a frequency at the Floquet frequency ($\omega_F \equiv \delta_\delta/2$). In this extended space, the harmonic components of the Schr\"{o}dinger eigenvalue equation obey a recursion relation:

\begin{widetext}
\begin{align}\label{eq:FLoquetRecursion}
    \begin{pmatrix}
        0 & 0 \\
        0 & -\Sigma_{\delta}/2 
    \end{pmatrix}
    \begin{pmatrix}
        a_N\\
        b_N
    \end{pmatrix} + 
    \begin{pmatrix}
        0 & \Omega_1/2 \\
        \Omega_2^*/2 & 0
    \end{pmatrix}  \begin{pmatrix}
        a_{N+1}\\
        b_{N+1}
    \end{pmatrix} +
        \begin{pmatrix}
        0 & \Omega_2/2 \\
        \Omega_1^*/2 & 0
    \end{pmatrix}  \begin{pmatrix}
        a_{N-1}\\
        b_{N-1}
    \end{pmatrix} = \left(\epsilon + N\hbar\omega_F\right)   \begin{pmatrix}
        a_N\\
        b_N
    \end{pmatrix}
\end{align}
\end{widetext}

In the dressed state basis, and for equal Rabi frequencies $|\Omega_1| = |\Omega_2| = |\Omega|$, the ladder of states separated by the Floquet frequency $\omega_F$ has a tri-diagonal representation that leads to 
a Bessel function representation $d_N\propto J_N\left(\Omega/\omega_F\right)$ within a dressed state manifold, where $d_N$ are the expansion coefficients of the atom-field coupled dressed states in the bare state basis \cite{van_leeuwen_autler-townes_1996}.

Using Eq.~\ref{eq:FLoquetRecursion} we may build an infinite-dimensional time-independent Floquet Hamiltonian $\mathcal{H}_F$ that satisfies the eigenvalue equation $\mathcal{H}_F\boldsymbol{\varphi}_N = \epsilon_N \boldsymbol{\varphi}_N$.
In practice, the infinite-dimensional Hamiltonian is truncated to some large value $N_{max}$ for which the solutions converge. 
All two-level models in this work use $N_{max}=50$, which was more than sufficient for reasonable convergence.
An example cutoff criterion is that the population in the $N_{max}^{\text{th}}$ sideband is small, i.e. $<0.1\%$ of the population. 

We obtain the eigenvectors and eigenenergies by diagonalizing the Floquet Hamiltonian $\mathcal{H}_F$. These eigenvectors and eigenenergies are the  Floquet modes and quasi-energies $\epsilon_N$, represented in the expanded Floquet basis. To compute the Floquet mode occupation given a specific input state, we project the initial state into the Floquet basis and compute the square of these amplitudes \cite{shirley_solution_1965}. We plot the mode occupation against quasienergy $\epsilon$ to generate the theory waterfall plots. The Floquet modes and quasienergies are sensitive to the detunings and the individual Rabi frequencies of the applied RF fields. In this work, we examine the spectral signatures of the Rydberg response as these parameters are varied.

\section{Experimental Results: Symmetric Detuning and Power-Balanced Fields} \label{sec:experimentalresults}

\begin{figure*}[htb]
    \centering     
    \begin{overpic}[width = \textwidth, tics=5, unit=1pt, grid = false, trim=0 0 0 0, clip]
    {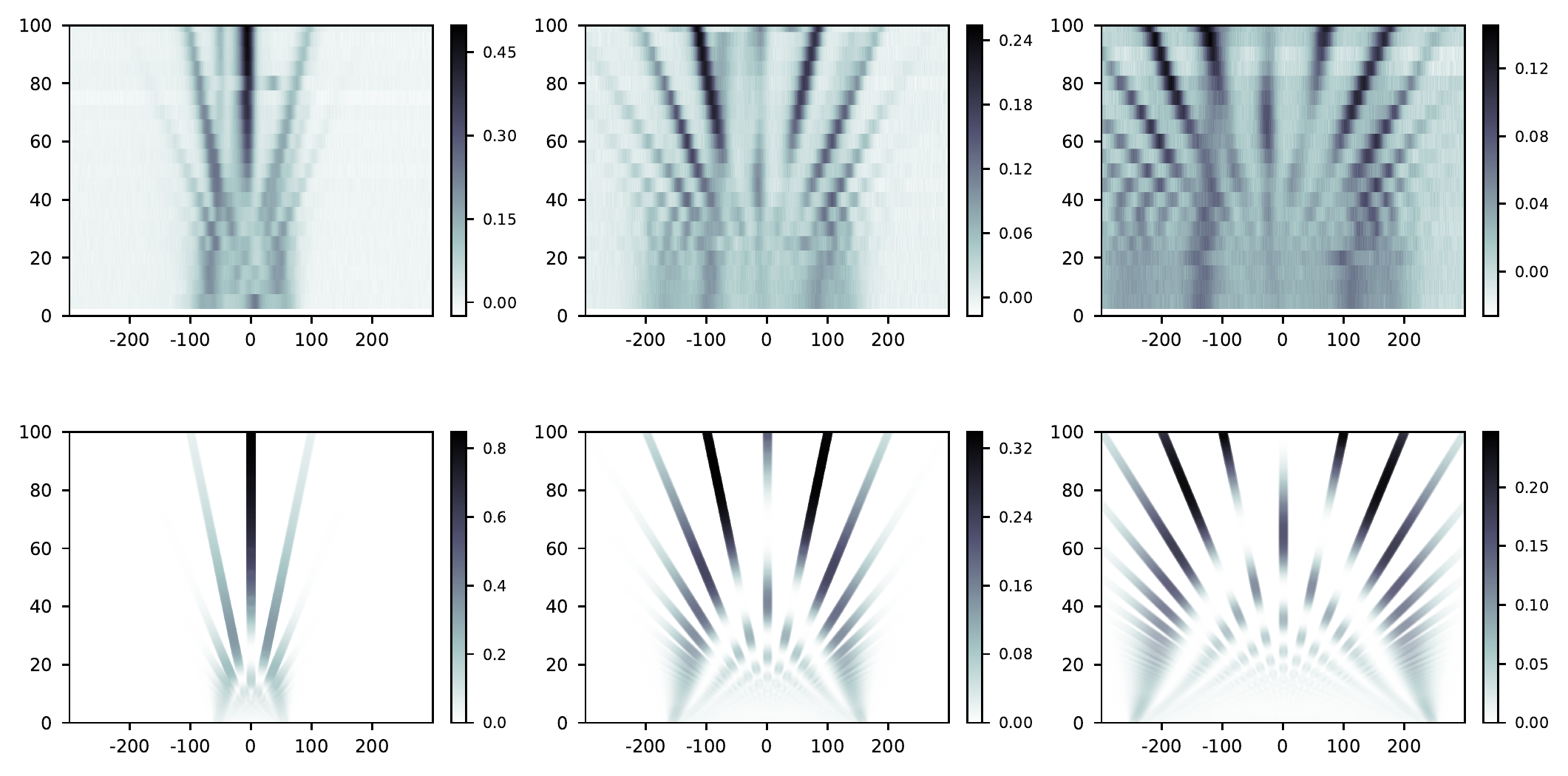}
    \put(23, 31){ \textcolor{black}{(a1)}   }
    \put(56, 31){ \textcolor{black}{(b1)}   }
    \put(89, 31){ \textcolor{black}{(c1)}   }
    \put(23, 5){ \textcolor{black}{(a2)}   }
    \put(56, 5){ \textcolor{black}{(b2)}   }
    \put(89, 5){ \textcolor{black}{(c2)}   }

    \put(16, 25.8){\makebox[0pt]{\footnotesize{ \Centerstack{\textcolor{black}{$\delta_c/2\pi$(MHz)}   }}}}
    \put(49, 25.8){\makebox[0pt]{ \footnotesize{ \Centerstack{\textcolor{black}{$\delta_c/2\pi$(MHz)}   }}}}
    \put(82, 25.8){\makebox[0pt]{ \footnotesize{ \Centerstack{\textcolor{black}{$\delta_c/2\pi$(MHz)}   }}}}

    \put(16, -0.3){\makebox[0pt]{ \footnotesize{ \Centerstack{\textcolor{black}{$\epsilon/2\pi$(MHz)}   }}}}
    \put(49, -0.3){\makebox[0pt]{ \footnotesize{ \Centerstack{\textcolor{black}{$\epsilon/2\pi$(MHz)}   }}}}
    \put(82, -0.3){\makebox[0pt]{ \footnotesize{ \Centerstack{\textcolor{black}{$\epsilon/2\pi$(MHz)}   }}}}

    \put(16, 49){\makebox[0pt]{ \Centerstack{\textcolor{black}{$\Omega/2\pi = 0.88\,\sqrt{\text{mW}}$}   }}}
    \put(49, 49){\makebox[0pt]{ \Centerstack{\textcolor{black}{$\Omega/2\pi = 2.02\,\sqrt{\text{mW}}$}   }}}
    \put(82, 49){\makebox[0pt]{ \Centerstack{\textcolor{black}{$\Omega/2\pi = 3.02\,\sqrt{\text{mW}}$}   }}}

    \put(16, 22.8){\makebox[0pt]{ \Centerstack{\textcolor{black}{$\Omega/2\pi = 57\,\text{MHz}$}   }}}
    \put(49, 22.8){\makebox[0pt]{ \Centerstack{\textcolor{black}{$\Omega/2\pi = 160\,\text{MHz}$}   }}}
    \put(82, 22.8){\makebox[0pt]{ \Centerstack{\textcolor{black}{$\Omega/2\pi = 250\,\text{MHz}$}   }}}
    
    \put(0, 38){\makebox[0pt]{\Centerstack{\rotatebox{90}{Exp. $\delta/2\pi$ (MHz)}}}}
    \put(0, 12.5){\makebox[0pt]{\Centerstack{\rotatebox{90}{Theory $\delta/2\pi$ (MHz)}}}}
        
    \end{overpic}
    \caption{Experimental (a1--c1) and theoretical (a2--c2) waterfall plots showing a simultaneous scan of $\delta_1 =-\delta_2 = \delta$, while the two Rabi frequencies $|\Omega_1| = |\Omega_2| = \Omega$ are kept constant. The mode quasi-energies are shown to increase with $\delta$ in each waterfall plot, and the splitting between the Floquet modes is linear in $\delta$ for the range of $\Omega$ shown. Increasing Rabi frequency has the effect of populating Floquet modes of higher Floquet photon index. At higher driving Rabi frequencies we see an overall shift of the spectra towards lower energy (as seen in (c1)). We attribute this shift to imperfect shielding of our reference cell from the applied RF fields.}
    \label{figure:3}
\end{figure*}

\begin{figure}
    \centering     
    \begin{overpic}[width = \columnwidth, tics=5, unit=1pt, grid = false, trim=0 0 0 0, clip]
    {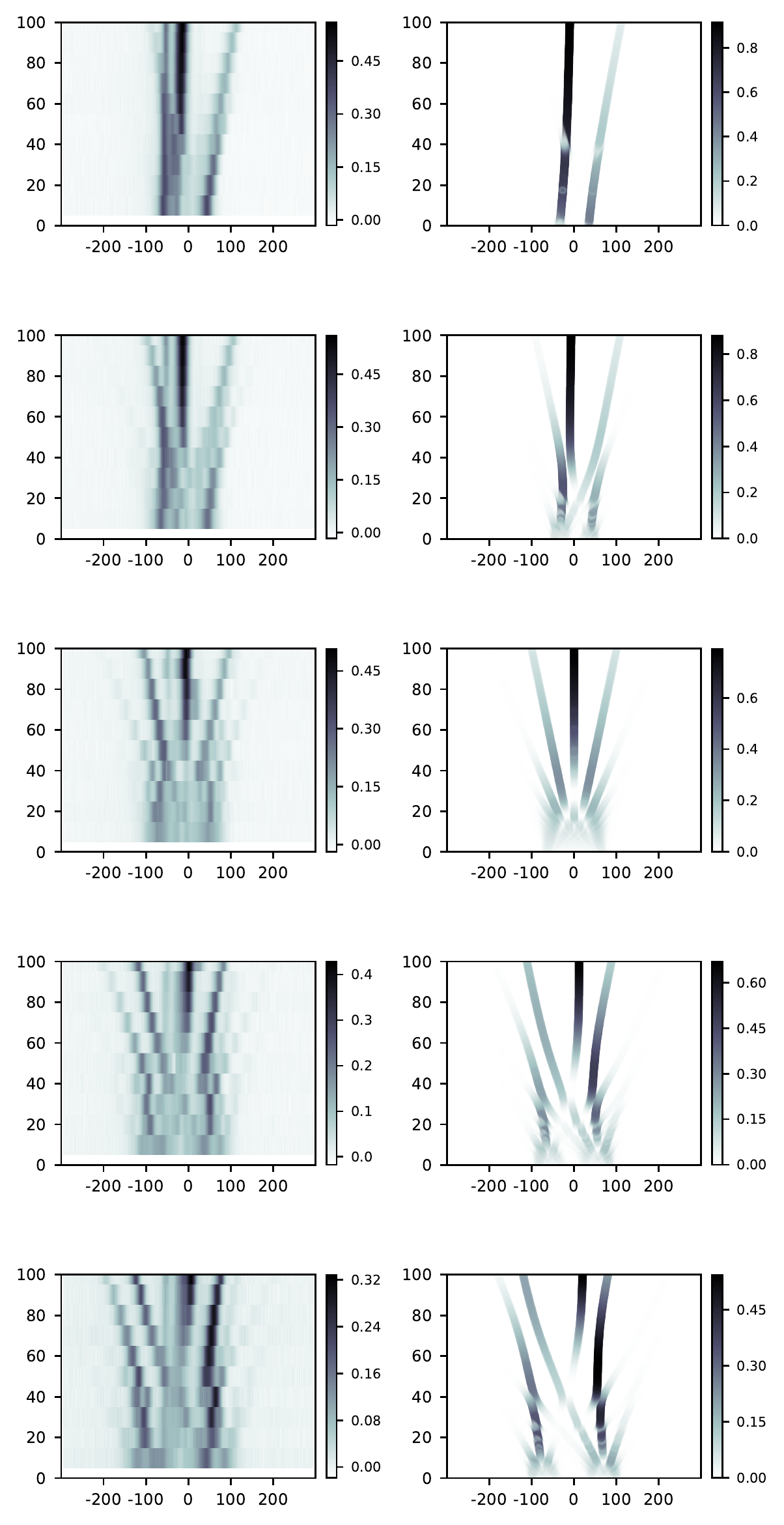}
    \put(16, 87){ \textcolor{black}{(a1)}   }
    \put(16, 66){ \textcolor{black}{(b1)}   }
    \put(16, 45){ \textcolor{black}{(c1)}   }
    \put(16, 25){ \textcolor{black}{(d1)}   }
    \put(16, 4.5){ \textcolor{black}{(e1)}   }
    \put(41, 87){ \textcolor{black}{(a2)}   }
    \put(41, 66){ \textcolor{black}{(b2)}   }
    \put(41, 45){ \textcolor{black}{(c2)}   }
    \put(41, 25){ \textcolor{black}{(d2)}   }
    \put(41, 4.5){ \textcolor{black}{(e2)}   }
    
    \put(12, 81.5){\makebox[0pt]{ \footnotesize{ \Centerstack{\textcolor{black}{$\delta_c/2\pi$(MHz)}   }}}}
    \put(12, 61){\makebox[0pt]{ \footnotesize{ \Centerstack{\textcolor{black}{$\delta_c/2\pi$(MHz)}   }}}}
    \put(12, 40.5){\makebox[0pt]{\footnotesize{ \Centerstack{\textcolor{black}{$\delta_c/2\pi$(MHz)}   }}}}
    \put(12, 20){\makebox[0pt]{\footnotesize{ \Centerstack{\textcolor{black}{$\delta_c/2\pi$(MHz)}   }}}}
    \put(12, -0.8){\makebox[0pt]{\footnotesize{ \Centerstack{\textcolor{black}{$\delta_c/2\pi$(MHz)}   }}}}
        
    \put(37, 81.5){\makebox[0pt]{ \footnotesize{ \Centerstack{\textcolor{black}{$\epsilon/2\pi$(MHz)}   }}}}
    \put(37, 61){\makebox[0pt]{ \footnotesize{ \Centerstack{\textcolor{black}{$\epsilon/2\pi$(MHz)}   }}}}
    \put(37, 40.5){\makebox[0pt]{ \footnotesize{ \Centerstack{\textcolor{black}{$\epsilon/2\pi$(MHz)}   }}}}
    \put(37, 20){\makebox[0pt]{ \footnotesize{ \Centerstack{\textcolor{black}{$\epsilon/2\pi$(MHz)}   }}}}
    \put(37, -0.8){\makebox[0pt]{ \footnotesize{ \Centerstack{\textcolor{black}{$\epsilon/2\pi$(MHz)}   }}}}

    \put(12, 99.5){\makebox[0pt]{ \Centerstack{\textcolor{black}{$\Omega_1/2\pi = 0.32\,\sqrt{\text{mW}}$}   }}}
    \put(12, 79){\makebox[0pt]{ \Centerstack{\textcolor{black}{$\Omega_1/2\pi = 0.61\,\sqrt{\text{mW}}$}   }}}
    \put(12, 58.5){\makebox[0pt]{ \Centerstack{\textcolor{black}{$\Omega_1/2\pi = 1\,\sqrt{\text{mW}}$}   }}}
    \put(12, 38){\makebox[0pt]{ \Centerstack{\textcolor{black}{$\Omega_1/2\pi = 1.44\,\sqrt{\text{mW}}$}   }}}
    \put(12, 17.5){\makebox[0pt]{ \Centerstack{\textcolor{black}{$\Omega_1/2\pi = 1.78\,\sqrt{\text{mW}}$}   }}}

    \put(37, 99.5){\makebox[0pt]{ \Centerstack{\textcolor{black}{$\Omega_1/2\pi = 6\,\text{MHz}$}   }}}
    \put(37, 79){\makebox[0pt]{ \Centerstack{\textcolor{black}{$\Omega_1/2\pi = 32\,\text{MHz}$}   }}}
    \put(37, 58.5){\makebox[0pt]{ \Centerstack{\textcolor{black}{$\Omega_1/2\pi = 67\,\text{MHz}$}   }}}
    \put(37, 38){\makebox[0pt]{ \Centerstack{\textcolor{black}{$\Omega_1/2\pi = 107\,\text{MHz}$}   }}}
    \put(37, 17.5){\makebox[0pt]{ \Centerstack{\textcolor{black}{$\Omega_1/2\pi = 138\,\text{MHz}$}   }}}
    \put(0, 10){\makebox[0pt]{\Centerstack{\rotatebox{90}{$\delta/2\pi$ (MHz)}}}}
    \put(0, 30){\makebox[0pt]{\Centerstack{\rotatebox{90}{$\delta/2\pi$ (MHz)}}}}
    \put(0, 51){\makebox[0pt]{\Centerstack{\rotatebox{90}{$\delta/2\pi$ (MHz)}}}}
    \put(0, 71){\makebox[0pt]{\Centerstack{\rotatebox{90}{$\delta/2\pi$ (MHz)}}}}
    \put(0, 91){\makebox[0pt]{\Centerstack{\rotatebox{90}{$\delta/2\pi$ (MHz)}}}}
    \end{overpic}
    \caption{Experimental (a1--e1) and theoretical (a2--e2) waterfall plots show a simultaneous scan of $\delta_1 =-\delta_2 = \delta$, while the two Rabi frequencies $|\Omega_1|$, as labeled, and $|\Omega_2| = 1\sqrt{\text{mW}}$ are kept constant. The imbalance in Rabi frequencies is apparent from the asymmetric spectra, and the sense of the imbalance shifts as the Rabi frequency $|\Omega_2|$ is larger or smaller than $|\Omega_2|$.}
    \label{figure:3_mis}
\end{figure}

\begin{figure*}[t]
    \centering
    \begin{overpic}[width = \textwidth, tics=5, grid = false, unit=1pt, trim=0 0 0 5, clip]
    {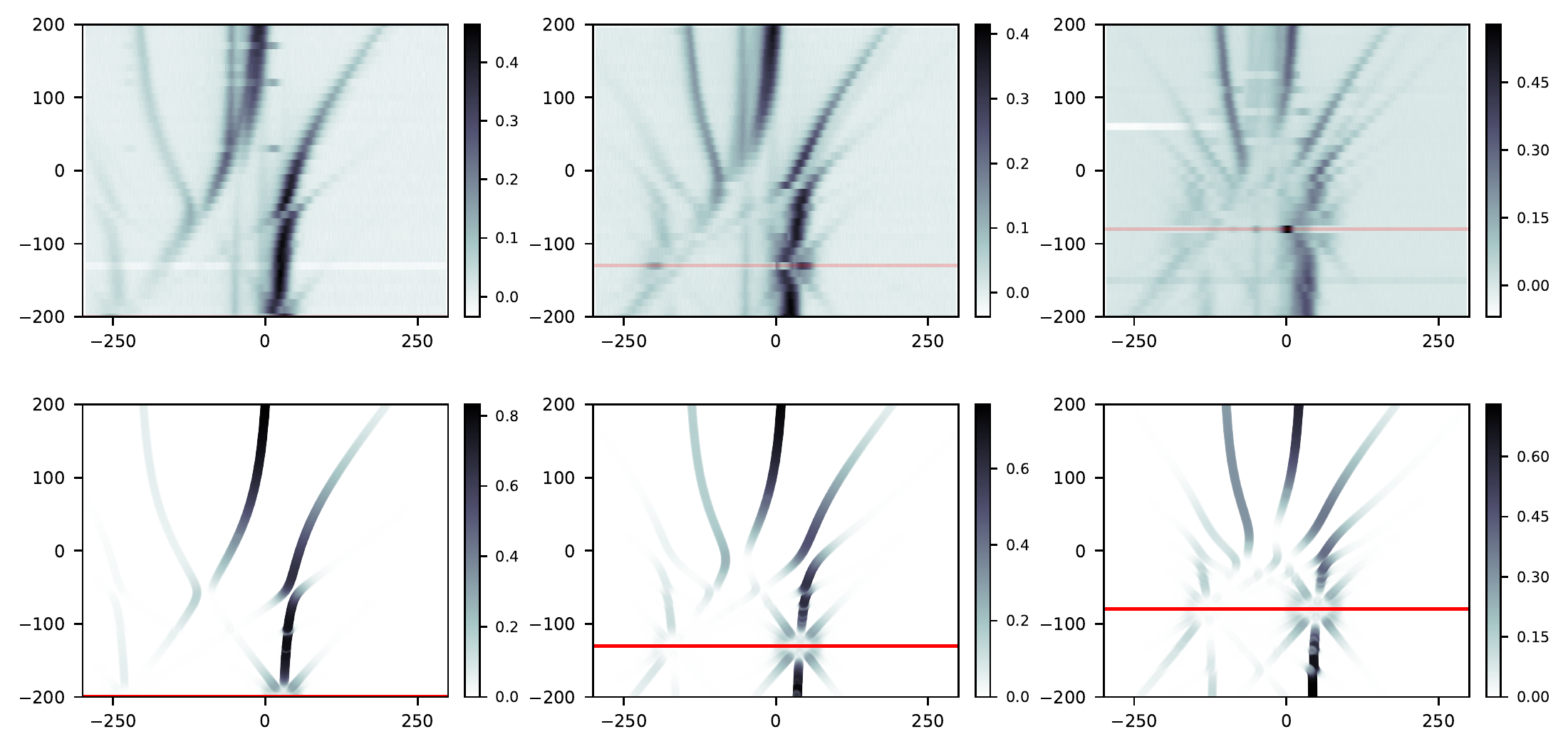}
        \put(24, 29){ \textcolor{black}{(a1)}   }
        \put(56, 29){ \textcolor{black}{(b1)}   }
        \put(89, 29){ \textcolor{black}{(c1)}   }
        
        \put(24, 4.5){ \textcolor{black}{(a2)}   }
        \put(56, 4.5){ \textcolor{black}{(b2)}   }
        \put(89, 4.5){ \textcolor{black}{(c2)}   }

        \put(17, 48){\makebox[0pt]{ \Centerstack{{\textcolor{black}{$\delta_2/2\pi = -200\,\text{(MHz)}$   }}}}}
        \put(50, 48){\makebox[0pt]{ \Centerstack{\textcolor{black}{$\delta_2/2\pi = -130\,\text{(MHz)}$   }}}}
        \put(83, 48){\makebox[0pt]{ \Centerstack{\textcolor{black}{$\delta_2/2\pi = -80\,\text{(MHz)}$   }}}}
        \put(17,23.8){\makebox[0pt]{ \footnotesize{ \Centerstack{\textcolor{black}{$\delta_c/2\pi$(MHz)}   }}}}
        \put(50,23.8){\makebox[0pt]{ \footnotesize{ \Centerstack{\textcolor{black}{$\delta_c/2\pi$(MHz)}   }}}}
        \put(83,23.8){\makebox[0pt]{ \footnotesize{ \Centerstack{\textcolor{black}{$\delta_c/2\pi$(MHz)}   }}}}
        \put(17,-0.8){\makebox[0pt]{ \footnotesize{ \Centerstack{\textcolor{black}{$\epsilon/2\pi$(MHz)}   }}}}
        \put(50,-0.8){\makebox[0pt]{ \footnotesize{ \Centerstack{\textcolor{black}{$\epsilon/2\pi$(MHz)}   }}}}
        \put(83,-0.8){\makebox[0pt]{ \footnotesize{ \Centerstack{\textcolor{black}{$\epsilon/2\pi$(MHz)}   }}}}
        \put(0, 37){\makebox[0pt]{\Centerstack{\rotatebox{90}{Exp. $\delta_1/2\pi$ (MHz)}}}}
        \put(0, 12.5){\makebox[0pt]{\Centerstack{\rotatebox{90}{Theory $\delta_1/2\pi$ (MHz)}}}}
    \end{overpic}\\
    \vspace{0.8cm}

    \begin{overpic}[width = 0.69\textwidth, tics=5, grid = false, unit=1pt, trim=0 0 0 5, clip]{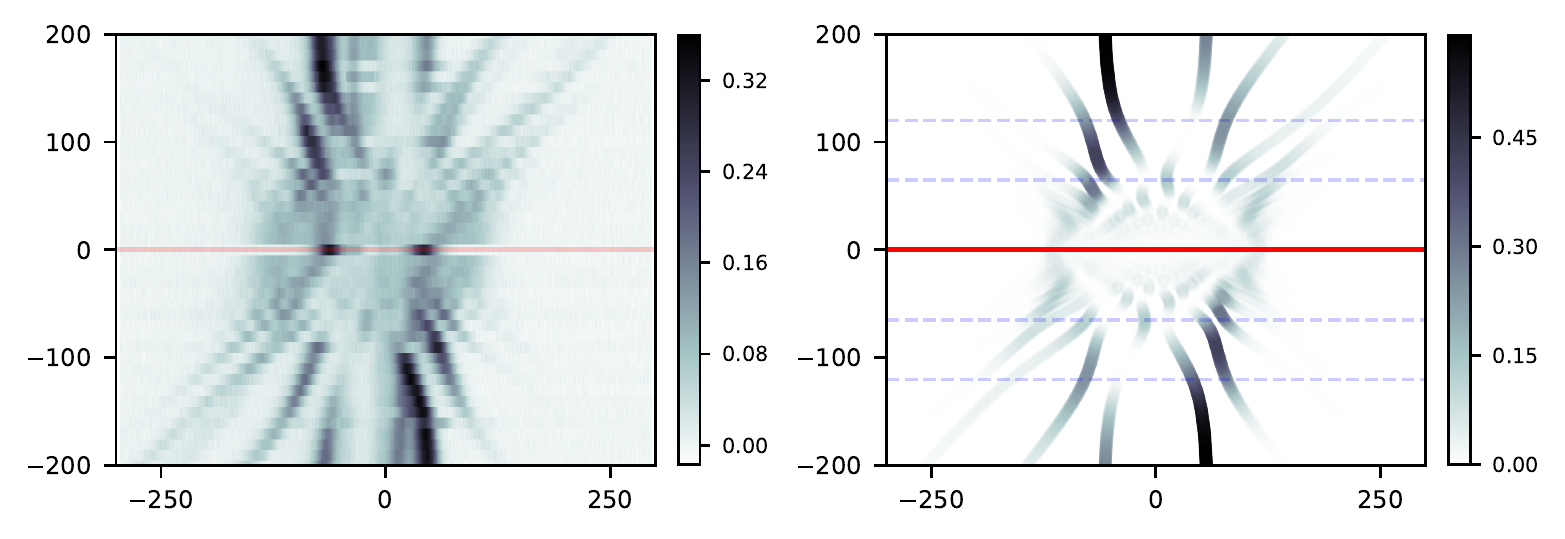}
        \put(35, 6.5){ \textcolor{black}{(d1)}   }     
        \put(84, 6.5){ \textcolor{black}{(d2)}   }
        
        \put(25, 35){\makebox[0pt]{ \Centerstack{\textcolor{black}{$\delta_1/2\pi = 0\,\text{(MHz)}$   }}}}
        \put(74, 35){\makebox[0pt]{ \Centerstack{\textcolor{black}{$\delta_2/2\pi = 0\,\text{(MHz)}$   }}}}
        \put(23, -0.8){\makebox[0pt]{ \Centerstack{\textcolor{black}{$\delta_c/2\pi$(MHz)}   }}}
        \put(74,-0.8){\makebox[0pt]{ \Centerstack{\textcolor{black}{$\epsilon/2\pi$(MHz)}   }}}
        \put(65.5, 26){\makebox[0pt]{ \Centerstack{\large\textcolor{black}{$\longrightarrow$}}}}
        \put(66, 22){\makebox[0pt]{ \Centerstack{\large\textcolor{black}{$\longrightarrow$}
        }}}
        \put(0, 18){\makebox[0pt]{\Centerstack{\rotatebox{90}{Exp. $\delta_1/2\pi$ (MHz)}}}}
        \put(102, 18){\makebox[0pt]{\Centerstack{\rotatebox{90}{Theory $\delta_1/2\pi$ (MHz)}}}}
    \end{overpic}\\
    \vspace{0.8cm}  
    
    \begin{overpic}[width = \textwidth, tics=5, grid = false, unit=1pt, trim=0 0 0 0, clip]{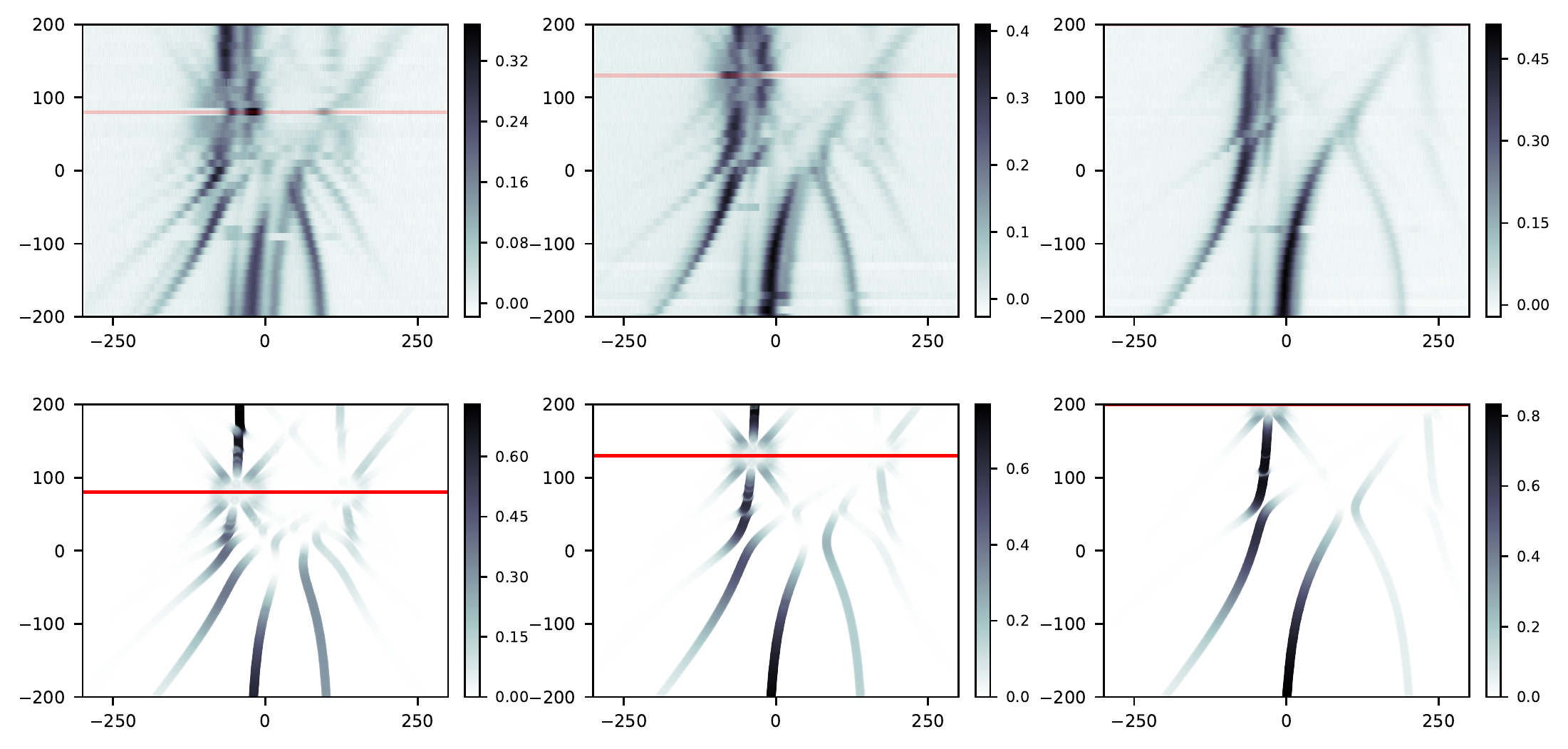}
        \put(24, 29){ \textcolor{black}{(e1)}   }
        \put(57, 29){ \textcolor{black}{(f1)}   }
        \put(89, 29){ \textcolor{black}{(g1)}   }
        
        \put(24, 4.5){ \textcolor{black}{(e2)}   }
        \put(57, 4.5){ \textcolor{black}{(f2)}   }
        \put(89, 4.5){ \textcolor{black}{(g2)}   }

        \put(17, 48){\makebox[0pt]{ \Centerstack{{\textcolor{black}{$\delta_2/2\pi = 80\,\text{(MHz)}$   }}}}}
        \put(50, 48){\makebox[0pt]{ \Centerstack{\textcolor{black}{$\delta_2/2\pi = 130\,\text{(MHz)}$   }}}}
        \put(83, 48){\makebox[0pt]{ \Centerstack{\textcolor{black}{$\delta_2/2\pi = 200\,\text{(MHz)}$   }}}}
        
        \put(17,23.8){\makebox[0pt]{ \Centerstack{\textcolor{black}{$\delta_c/2\pi$(MHz)}   }}}
        \put(50,23.8){\makebox[0pt]{ \Centerstack{\textcolor{black}{$\delta_c/2\pi$(MHz)}   }}}
        \put(83,23.8){\makebox[0pt]{ \Centerstack{\textcolor{black}{$\delta_c/2\pi$(MHz)}   }}}

        \put(17,-0.8){\makebox[0pt]{ \Centerstack{\textcolor{black}{$\epsilon/2\pi$(MHz)}   }}}
        \put(50,-0.8){\makebox[0pt]{ \Centerstack{\textcolor{black}{$\epsilon/2\pi$(MHz)}   }}}
        \put(83,-0.8){\makebox[0pt]{ \Centerstack{\textcolor{black}{$\epsilon/2\pi$(MHz)}   }}}

        \put(0, 37){\makebox[0pt]{\Centerstack{\rotatebox{90}{Exp. $\delta_1/2\pi$ (MHz)}}}}
        \put(0, 12.5){\makebox[0pt]{\Centerstack{\rotatebox{90}{Theory $\delta_1/2\pi$ (MHz)}}}}
    \end{overpic}
    \caption{
   (a1--g1): Experimental waterfall plots show the obtained spectra against coupling laser frequency ($\delta_c$) as the detuning $\delta_1$ is scanned for constant $\delta_2$ and Rabi frequencies $|\Omega_1| = |\Omega_2| = 105$~MHz. (a2--g2): Numerically obtained Floquet quasienergy spectra showing Floquet mode occupations against quasienergies ($\epsilon$) as $\delta_1$ is scanned. The overlaid red lines indicate equal detunings of the two fields ($\delta_1 = \delta_2$). 
    The arrows and overlaid dashed blue lines in (d') indicate subharmonic resonances.}
    \label{figure:4}
\end{figure*}

Our main results are the observed agreement between experimental Rydberg EIT spectra and a theoretical model of the dual-tone driven Rydberg system. 
We predict and observe many nonlinear and multiphoton effects due to the dual-tone RF dressing. 

We present experiment alongside theoretical predictions, using similar axes with different units, matching experimental and theory parameters. 
The energy spectrum is scanned experimentally by laser detuning $\delta_c$, and state theory energy is given in quasienergy $\epsilon$. 
The color/darkness axis represents experimental transmittance (EIT), alongside theory population projections of the $61D_{5/2}$ state into the Floquet modes. 
The waterfall axis scans applied field strength in $\sqrt{\text{mW}}$, pairing with Rabi frequencies $\Omega$. We convert between Rabi frequency and applied field with a two-point function using Rabi frequencies of $2\pi\cdot\{70, 160\}$~MHz for applied RF powers of $\{0.24, 6.12\}$~dBm, which agrees well with the experimental data, and corresponds to a linear region of a single-field Autler--Townes scan of Rabi frequencies. We compensate the powers sent to the horn to maintain field strength ($\sqrt{\textrm{mW}}$ or $|\Omega|$) across the horn's frequency-dependant gain curve.

In the following sections we discuss the spectral features of experimentally obtained EIT spectra for a variety of configurations of RF fields. The main features of our experimental spectra are very well modeled by a simple two-state Hamiltonian that includes just the Rydberg states $61D_{5/2}$ and $62P_{3/2}$, coupled with a linearly polarized dual-tone RF field. 
In this Section, we demonstrate  ``symmetric" RF dual-tone fields that are symmetrically detuned ($\delta_1 = -\delta_2 = \delta$) from the Rydberg resonance frequency of $9.226$~GHz, and power balanced, with equal Rabi frequencies ($|\Omega_1| = |\Omega_2| = |\Omega|$ ).
In Sec.~\ref{sec:symmetric-fields-scan-rabi}, we scan Rabi frequency $\Omega$ for a few fixed detunings $\delta$, and in Sec.~\ref{sec:scan-detuning}, we scan detuning $\delta$ for a few fixed Rabi frequencies $\Omega$. 

\subsubsection{Scanning Rabi frequencies}\label{sec:symmetric-fields-scan-rabi}
Our first set of results is presented in Fig.~\ref{figure:2}, which shows waterfall plots of Rydberg EIT spectra as the symmetric detuning $\delta$ is held constant while both power-balanced Rabi frequencies $|\Omega|$ are simultaneously scanned. We present data for three different values of $\delta$.

For this case, the two-state model predicts equal and opposite energy shifts of the states at each Rabi frequency; the eigenvalues of the Floquet Hamiltonian remain identically the diagonal elements. The quasienergies therefore remain unchanged as the Rabi frequency is scanned, resulting in spectra where the spacing of the Floquet modes is set entirely by the detunings, and this mode spacing remains constant as Rabi frequencies are swept. 

The Floquet mode occupation is, however, sensitive to the Rabi frequencies: higher Rabi frequencies populate modes with higher photon number. The mode occupations follow the Bessel functions in this two state model (the inclusion of more states causes deviations from this Bessel function behaviour): $d_N \propto J_{N}\left(\Omega/\delta\right)$, where $N$ signifies the Floquet mode index, and $\Omega$ and $\delta$ are the balanced Rabi frequencies and symmetric detunings of the two RF tones \cite{van_leeuwen_autler-townes_1996}. The Bessel functions are visually represented in the data of Fig.~\ref{figure:2}, as seen by following the peak heights corresponding to a specific value of detuning $\delta$ as we scan the dressing field power (that is, looking at a vertical slice of a waterfall plot at the locations $\delta_c = N\delta$).

We note that the original investigations of bichromatic optical driving of two-level atoms showed distinct linewidths for the odd and even Floquet modes; these widths were later shown to depend on the ratio of the Rabi frequency to the Floquet frequency ($\delta_\delta/2$), as well as the natural widths of the atomic states \cite{van_leeuwen_autler-townes_1996}. In our data, the two Rydberg levels have very similar decay rates, and state decay linewidths in the kHz regime---the state lifetimes at room temperature, computed using Ref.~\cite{sibalic_arc_2017}, are $\tau_{61D_{5/2}} \approx 110\,\mu$s and $\tau_{62P_{3/2}} \approx 145\,\mu$s---and we observe no obvious difference in the linewidths of the even and odd peaks, which are primarily set by the Doppler-broadened EIT linewidth (typically several MHz).

\subsubsection{Scanning Detunings}\label{sec:scan-detuning}
Our second set of results is presented in Fig.~\ref{figure:3}, which shows waterfall plots of Rydberg EIT spectra as the power-balanced equal Rabi frequencies of the two fields are constant while the symmetric detunings $\delta_1 = -\delta_2 = \delta$ are simultaneously scanned. We present data for three different values of $\Omega$.
We note that the spectra are primarily visible between $-\Omega < \delta_c < +\Omega $.

Again, the symmetric detunings each produce equal and opposite energy shifts of the states, and the eigenvalues of the Floquet Hamiltonian are determined by its diagonal elements. These eigenvalues are thus spaced by multiples of the Floquet frequency $\omega_F\equiv\delta_{\delta}/2$. The mode spacing changes linearly over the waterfall plot with detuning---this is in contrast to Fig.~\ref{figure:2} where the constant detunings (labeling each plot) resulted in equal-spaced spectral features over the waterfall with Rabi frequency. 

Here we see that for the three different values of Rabi frequency the behaviour of the quasienergies with detuning remains the same. However, higher Rabi frequencies drive population to higher-order Floquet modes, resulting in a larger fan-out of the spectra; note the different scales of the colormaps for the three cases, as population is spread out among a greater number of Floquet modes. The Bessel functions determining the mode occupation, $d_N \propto J_{N}\left(\Omega/\delta\right)$, are less visually apparent in these spectra as the denominator of the Bessel function argument changes over the waterfall.

\section{Application Scenario: Asymmetric and Unbalanced Fields}\label{sec:applicationscenario}
With an eye toward application scenarios, we consider the case where one field parameter is held, and the other controlled field is swept in power or detuning, while observing the dual-tone Rydberg EIT spectrum. 
We note that any particular spectrum can give information about both tones, moreover if one can `tune' into the `symmetric' and `balanced' cases, the known field strength and frequency can give the `unknown' RF signal's parameters. 
We illustrate here trends which appear as one moves away in either `asymmetric' detuning or `unbalanced' power scenarios. 

In Sec \ref{sec:symmetric-fields-scan-rabi} we demonstrated the dependence of the symmetric dual-tone spectra on detuning and Rabi frequency. The quasienergies remain constant with Rabi frequency for symmetric detunings and unbalanced Rabi frequencies. In general, the spectra are more complex functions of these 
parameters. For instance, for asymmetric detunings or Rabi frequencies, the quasienergies oscillate as the symmetric Rabi frequencies are simultaneously scanned. The spectra of Fig.~\ref{figure:2} and Fig.~\ref{figure:3} offer a tool for characterizing an unknown RF frequency through frequency and power matching with scans of the Rabi frequencies and detuning. 

\subsubsection{Unbalanced Power with Symmetric Detuning}  Our third set of results is presented in Fig.~\ref{figure:3_mis}, which shows waterfall plots of Rydberg EIT spectra 
for unbalanced but constant Rabi frequencies as the symmetric detunings $\delta_1 = -\delta_2 = \delta$ are simultaneously scanned. In these plots, $|\Omega_2| = 1 \sqrt{\text{mW}}$ was held constant, and we present data for five different values of $|\Omega_1|$.

The mismatched Rabi frequencies cause an asymmetric spectrum where the shift is indicative of the relative Rabi frequencies. A similar behaviour can be observed for mismatched detunings. These asymmetric spectra thus allow a determination of the detuning of an unknown RF field based on the symmetry of the quasienergy curves.

\subsubsection{Asymmetric Detuning with Balanced Power} 
Our final set of results is presented in Fig.~\ref{figure:4}, which shows waterfall plots of Rydberg EIT spectra where the equal Rabi frequencies of the two fields and the detuning $\delta_2$ are held constant, while $\delta_1$ is swept through resonance. We present data for seven different values of $\delta_2$. 

This is a realistic scenario for instance in the case where spurious signals or jamming signals are present. More complex spectra arise in this case. We make several observations of the main features of such spectra in the two-field case, below.

First, when $\delta_2$ is far off resonance, the spectrum is essentially the Autler--Townes spectrum of a single field as it sweeps through the Rydberg resonance. The spectra are largely mirrored for the cases where the constant detuning is $\delta_2$ vs. $-\delta_2$, for example Fig~\ref{figure:4}~(a1) and~(g1).

Second, around the region of equal detunings (marked with red lines in Fig.~\ref{figure:4}), we observe a phase diffusion effect in a region where the two RF fields are within a linewidth. This is due to the two fields inherently acting as a field at a single frequency but modulated according to their relative random phase. The measured spectrum is then essentially a time-averaged spectrum with random relative phases that effectively produce a random amplitude modulation of the Rabi splitting.
    
Third, in Fig.~\ref{figure:4}\,(d1) and~(d2), we have the case where $\delta_1 =0$, that is, one field is on resonance with the Rydberg transition. The red lines thus indicate the case where both fields are on resonance. This case is interesting in that the data on either side of this red line (and with $\delta_2$ detuned up to several tens of kHz) corresponds to the atom-mixer configurations typically used to detect RF fields within a heterodyne configuration \cite{simons_rydberg_2019}. 

Avoided crossings appear at subharmonic resonances (marked with dashed blue lines in Fig.~\ref{figure:4} (d2)) of the Rabi frequency $\Omega$; when $\delta_1 = 0$ these are resonances at $\delta_2 = \pm\Omega/k$, with small shifts from these values caused by ac Stark shifts from the applied fields. In our spectra, these subharmonic resonances are further shifted due to the different generalized Rabi frequencies of our fields, since the field strengths were kept equal as the detuning of a single field was scanned.

\section{Discussion and Conclusion}\label{sec:conclusion}
In this work we have explored the EIT spectra of Rydberg atoms dressed with intense dual-tone RF fields. These spectra are qualitatively different from the Autler--Townes spectra of Rydberg atoms dressed with a single field. The large dipole moments of Rydberg atoms allow these effects to be investigated with modest RF powers. Our analysis validates the broad applicability of theoretical models that treat the two-level dynamics of the RF-dressed Rydberg levels, while the optical fields act as indirect probes of the RF-induced multiphoton dynamics. This will allow a simplified treatment of dual-tone and multi-tone RF dressing in the Rydberg system that circumvents computationally intensive steady-state calculations with the full system density matrix. Such steady-state calculations moreover, are non-trivial to perform in the case of dissipative systems subject to driving outside the high-frequency regime, where the interplay of driving and dissipation become important.

Dual-tone RF dressing of Rydberg states unlocks a new toolbox that we foresee has implications for novel detection schemes in low frequency RF field sensing. For instance, we may realise phase-sensitive detection of fields in the $100\,$MHz range by coupling the energy levels of the synthetic Floquet dimension with the low frequency field to be detected, creating a three-frequency loop scheme. Previous schemes for electric field detection in the low-frequency and dc regimes have employed similar Floquet sideband techniques \cite{rotunno2023detection,Miller_2016, Bason_2010}. However in those studies, the Stark shifting of energies caused field-dependent shifts of the atomic spectra, which necessitates careful calculation and re-tuning due to Stark shifts at the higher fields necessary to generate higher-order Floquet sidebands. In contrast, the symmetric shifts caused by symmetric dual-tone drives in the Rydberg system allows the observation of spectra in high Floquet modes but without the added complication of overall spectral shifts. 

Our work enhances our understanding of the behavior of the system in an amplitude modulated field; in general the symmetric dual-tone dressing may be described as a $100\%$ sine-wave amplitude modulated field. Amplitude modulation has been gainfully demonstrated for Rydberg-based antennas, where studies of bandwidth and sensitivity are at the forefront of research in the field. The investigations performed in this paper allow us to place on firmer footing our understanding of the response to amplitude modulation in the system. 

Our investigations of these complex spectra will also be important as Rydberg atom-based electrometry transitions to real-world applications, where simple AT spectra may be significantly distorted and complicated by spurious tones that may be present, for example in the case of electromagnetic signal jamming. A deeper understanding of the response of Rydberg systems to multiple tones with variable detunings and powers will be necessary in order to unravel complex real-world spectra and to obtain meaningful results.

\section*{Acknowledgements}
The authors would like to express their gratitude for conversations on further extensions of the theoretical model with R. M. Potvliege, E. L. Shirley, and S. Eckel.

\appendix

\begin{figure*}[htb]
    \centering  
    \begin{overpic}[width = \textwidth, tics=5, unit=1pt, grid = false, trim=0 0 0 0, clip]
    {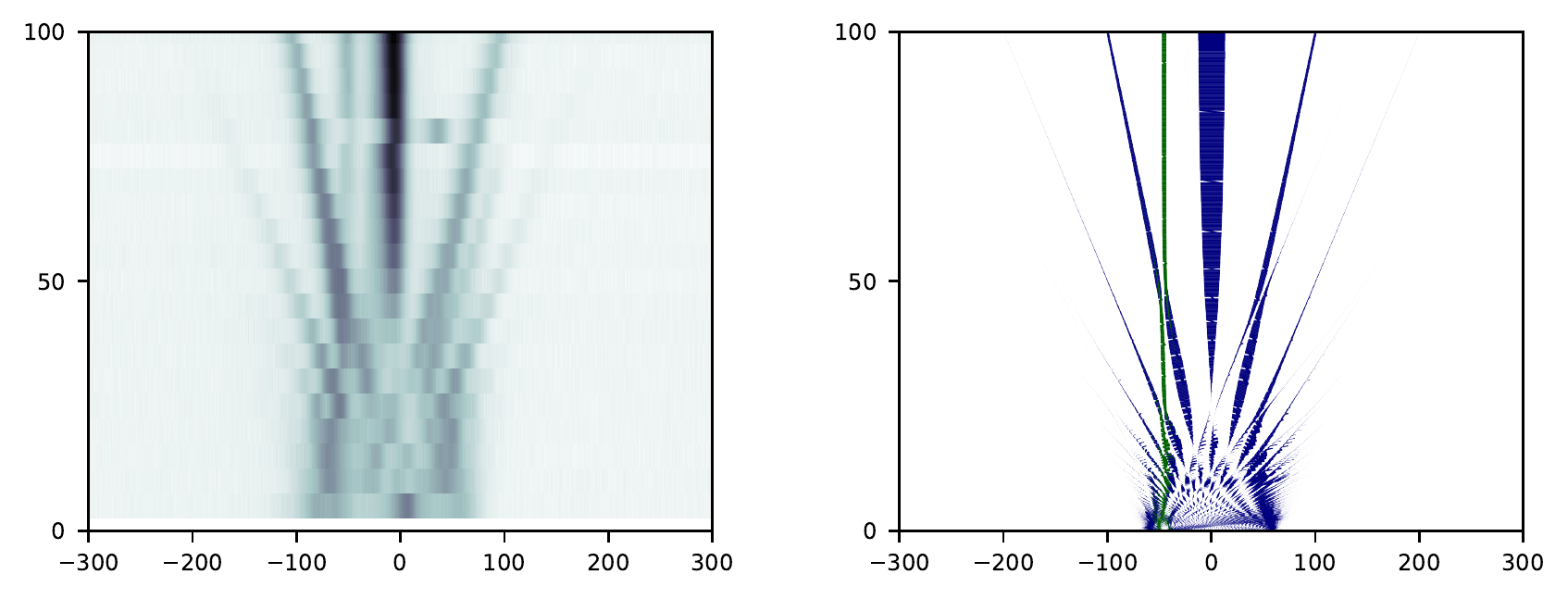}
    \put(42, 6){ \textcolor{black}{(a)}   }
    \put(93, 6){ \textcolor{black}{(b)}   }

    \put(16, 50){\makebox[0pt]{ \Centerstack{\textcolor{black}{}   }}}
    
    \put(27, -0.8){\makebox[0pt]{\footnotesize{ \Centerstack{\textcolor{black}{$\delta_c/2\pi$(MHz)}   }}}}
    \put(77, -0.8){\makebox[0pt]{ \footnotesize{ \Centerstack{\textcolor{black}{$\epsilon/2\pi$(MHz)}   }}}}
    \put(27, 37.5){\makebox[0pt]{\Centerstack{\rotatebox{0}{ $\Omega/2\pi = 0.88\,\sqrt{\text{mW}}$}}}}
    \put(77, 37.5){\makebox[0pt]{\Centerstack{\rotatebox{0}{ $\Omega/2\pi = 57\,\text{MHz}\quad\text{4-level model}$}}}}
    \put(0, 20){\makebox[0pt]{\Centerstack{\rotatebox{90}{Exp. $\delta/2\pi$ (MHz)}}}}
    \put(51, 20){\makebox[0pt]{\Centerstack{\rotatebox{90}{Theory $\delta/2\pi$ (MHz)}}}}
    \put(19.5, 32){\makebox[0pt]{ \Centerstack{\large\textcolor{Green}{$\Longrightarrow$}        }}}
    \put(24, 32){\makebox[0pt]{ \Centerstack{\large\textcolor{Green}
    {$\Longleftarrow$}        }}}
    \put(30, 16){\makebox[0pt]{ \Centerstack{\rotatebox{140}{\large\textcolor{blue}{$\Longrightarrow$}    }}}}
    \put(25, 20){\makebox[0pt]{ \Centerstack{\rotatebox{140}{\large\textcolor{blue}{$\Longleftarrow$}        }}}}
    \put(33.5, 25){\makebox[0pt]{ \Centerstack{\rotatebox{0}{\large\textcolor{black}{$\Longleftarrow$}    }}}}
    \end{overpic}
    
    \caption{The 4-state model: (a) The data of Fig.~\ref{figure:3}\,(a1) (symmetric, power-balanced fields as the detuning is scanned) shows structure that is not accounted for in the two-level model. (b) Including the fine states in the model reproduces these features. Here for visibility we represent the Floquet mode occupations as the widths of the spectral lines. The overlap of the Floquet quasienergies with the zero Floquet mode of $61 D_{5/2}$ is shown in blue, while the overlap with the zero Floquet mode of $61 D_{3/2}$ is shown in green. We scale the initial state occupations to reflect the relative peak heights of the $61 D_{3/2}$ and $61 D_{5/2}$ states in order to facilitate a comparison.}
    \label{figure:5}
\end{figure*}
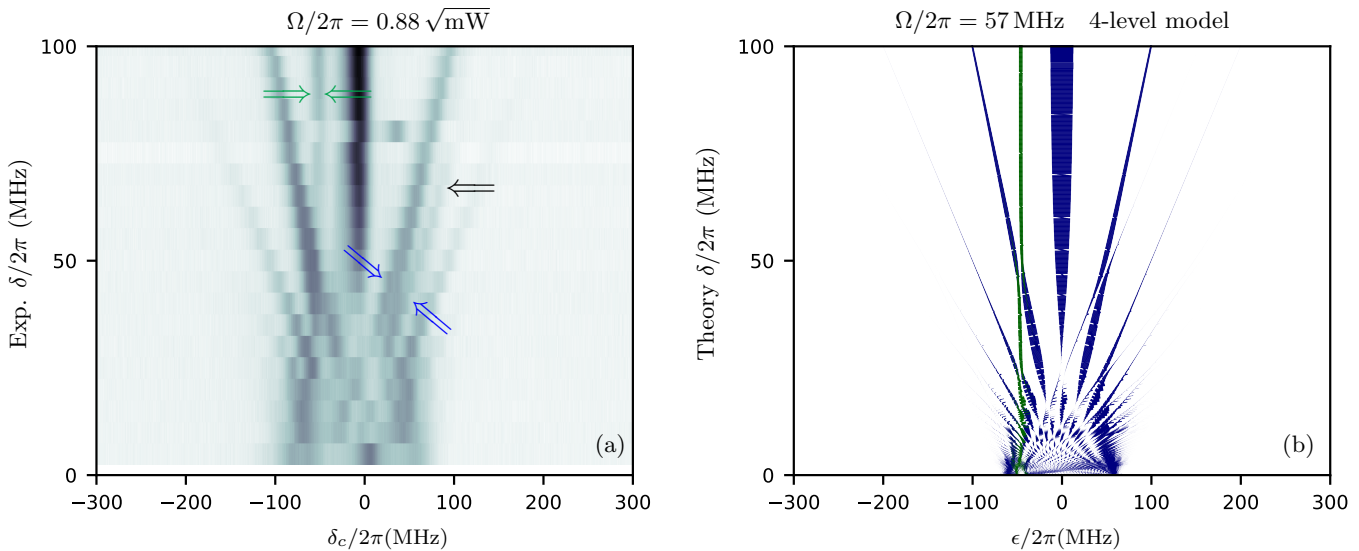

\section{Fine structure and magnetic sublevels}\label{sec:finestructure}

An examination of the data in Fig.~\ref{figure:3} and Fig.~\ref{figure:4} shows some spectral structure that is not accounted for in our two-level model. In this appendix we present extensions of the two-level analysis that include the fine-structure splitting of the two Rydberg states, and also the magnetic sublevels of the fine structure states.

\subsection{4-level model: Fine structure of Rydberg levels}
We show in Fig.~\ref{figure:5}\,(a) the data of Fig.~\ref{figure:3}\,(a1), emphasizing spectral features that are not reproduced in the two-level model. In Fig.~\ref{figure:5}\,(b) we show results from a four-state computation that includes the states $61D_{5/2}$, $61D_{3/2}$, $62P_{3/2}$, and $62P_{1/2}$ (see Appx.~\ref{sec:numbers}). The results reproduce some of the finer spectral structure that we observe in the data:

First, the green arrows in the figure highlight the appearance of the $61D_{3/2}$ fine-state to the left of the zero-detuning coupling laser resonance ($\delta_c = 0$) in the EIT spectra. This feature indicates an initial population in the $61D_{3/2}$ state that does not appear to participate significantly in the dynamics. We likewise note that the $61D_{5/2}$ population does not show significant mixing into this state. 

Second, the blue arrows in the figure show an apparent splitting of the spectral line due to an avoided crossing with the Floquet quasienergies associated with the fine structure. A similar avoided crossing appears on the left of the central mode.

Third, the black arrow highlights a spectral feature due to the fine-structure that persists and appears more clearly for the slightly higher Rabi frequency of Figure~\ref{figure:3}\,(b1): this feature is quite pronounced above the $N=2$ mode, and is absent above the $N=-2$ mode.

Fourth and lastly, an interesting feature that is not reproduced by our model is seen at quasienergies of $\approx-75$~MHz in Figure~\ref{figure:3}\,(b1). This is likely due to mixing from the hyperfine levels on the optical transition, which is expected to appear at this energy when Doppler mismatch is accounted for.

\begin{figure*}
    \centering
    \begin{overpic}[width = \textwidth, tics=5, unit=1pt, grid = false, trim=0 0 0 0, clip]
    {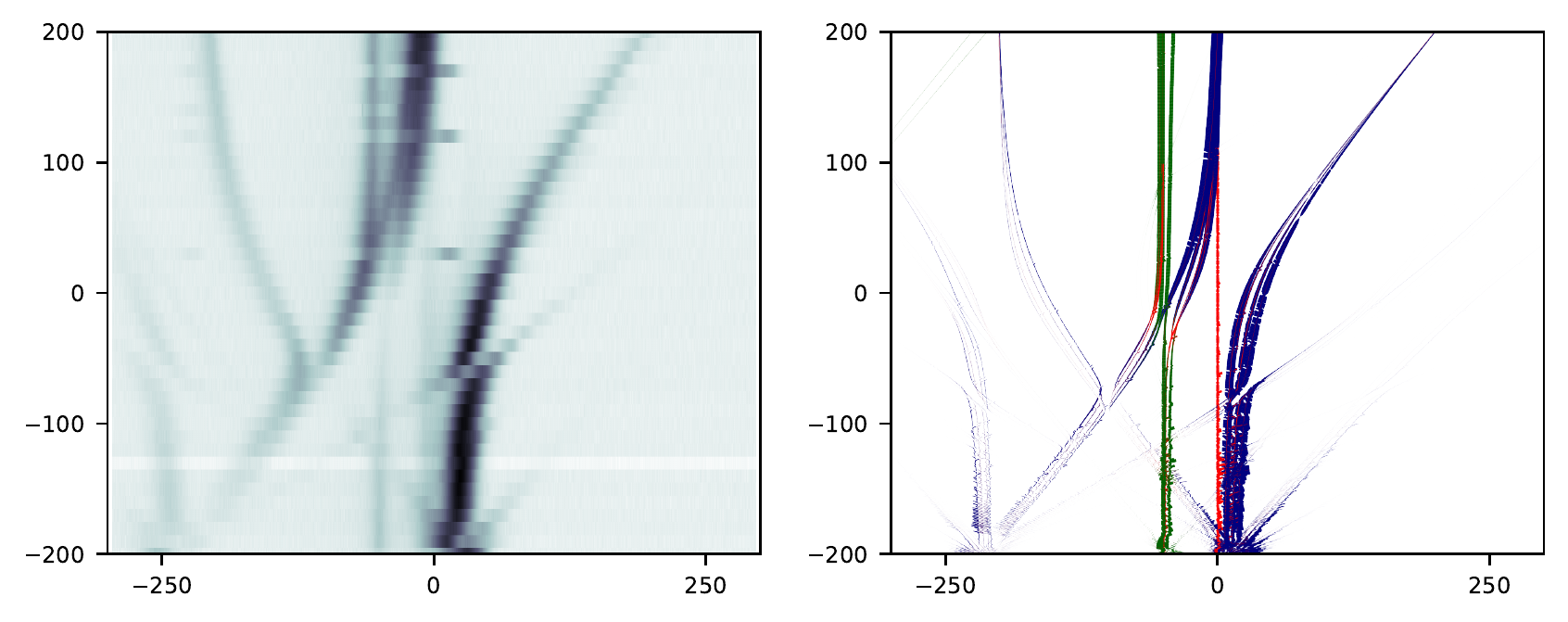}
    \put(42, 6){ \textcolor{black}{(a)}   }
    \put(93, 6){ \textcolor{black}{(b)}   }

    \put(27, -0.8){\makebox[0pt]{\footnotesize{ \Centerstack{\textcolor{black}{$\delta_c/2\pi$(MHz)}   }}}}
    \put(77, -0.8){\makebox[0pt]{ \footnotesize{ \Centerstack{\textcolor{black}{$\epsilon/2\pi$(MHz)}   }}}}
    \put(27, 39){\makebox[0pt]{\Centerstack{\rotatebox{0}{ $\delta_2/2\pi = 200\,$(MHz)}}}}
    \put(77, 39){\makebox[0pt]{\Centerstack{\rotatebox{0}{ $\delta_2/2\pi = 200\,$(MHz)$\quad\text{16-level model} $}}}}
    \put(0, 21){\makebox[0pt]{\Centerstack{\rotatebox{90}{Exp. $\delta_1/2\pi$ (MHz)}}}}
    \put(51, 21){\makebox[0pt]{\Centerstack{\rotatebox{90}{Theory $\delta_1/2\pi$ (MHz)}}}}
    \put(21, 5){\makebox[0pt]{ \Centerstack{\large\textcolor{Green}{$\Longrightarrow$}
    }}}
    \put(26, 5){\makebox[0pt]{ \Centerstack{\large\textcolor{Green}{$\Longleftarrow$}
        }}}
    \put(25, 16){\makebox[0pt]{ \Centerstack{\rotatebox{0}{\large\textcolor{red}{$\Longrightarrow$}    }}}}
    \put(29.5, 16){\makebox[0pt]{ \Centerstack{\rotatebox{0}{\large\textcolor{red}{$\Longleftarrow$}    }}}}
    \put(21, 27){\makebox[0pt]{ \Centerstack{\rotatebox{0}{\large\textcolor{black}{$\Longrightarrow$}    }}}}
    \put(29, 27){\makebox[0pt]{ \Centerstack{\rotatebox{0}{\large\textcolor{black}{$\Longleftarrow$}    }}}}
    \end{overpic}
    \caption{The 16-state model: (a) The data of Fig.~\ref{figure:4}\,(a1), with asymmetric detunings and power-balanced Rabi frequencies, as the detuning $\delta_1$ is scanned. (b) Including the magnetic sublevels $m_J$ of the fine states in the model reproduces more features from the data. Here for visibility we represent the Floquet mode occupations as the widths of the spectral lines. The overlap of the Floquet quasienergies with the $m_J = \pm 1/2$ sublevels of the zero Floquet mode of $61 D_{5/2}$ is shown in navy, the overlap with the $m_J = \pm 1/2$ sublevels of the zero Floquet mode of $61 D_{3/2}$ is shown in green, and other $m_J$ components ($|m_J|>1/2$) of both lines are shown in red.  We scale the initial state occupations of the $m_J = \pm 1/2$ components of $61 D_{3/2}$ and $61 D_{5/2}$  to reflect their relative peak heights from a simple EIT scan, and scale the initial state occupations of all the other $m_J$ sublevels to be $10\%$ of the $m_J = \pm 1/2$ occupation of $61 D_{5/2}$; in general these would depend on the polarization anisotropy and mixing of the magnetic sublevels.}
    \label{figure:6}
\end{figure*}
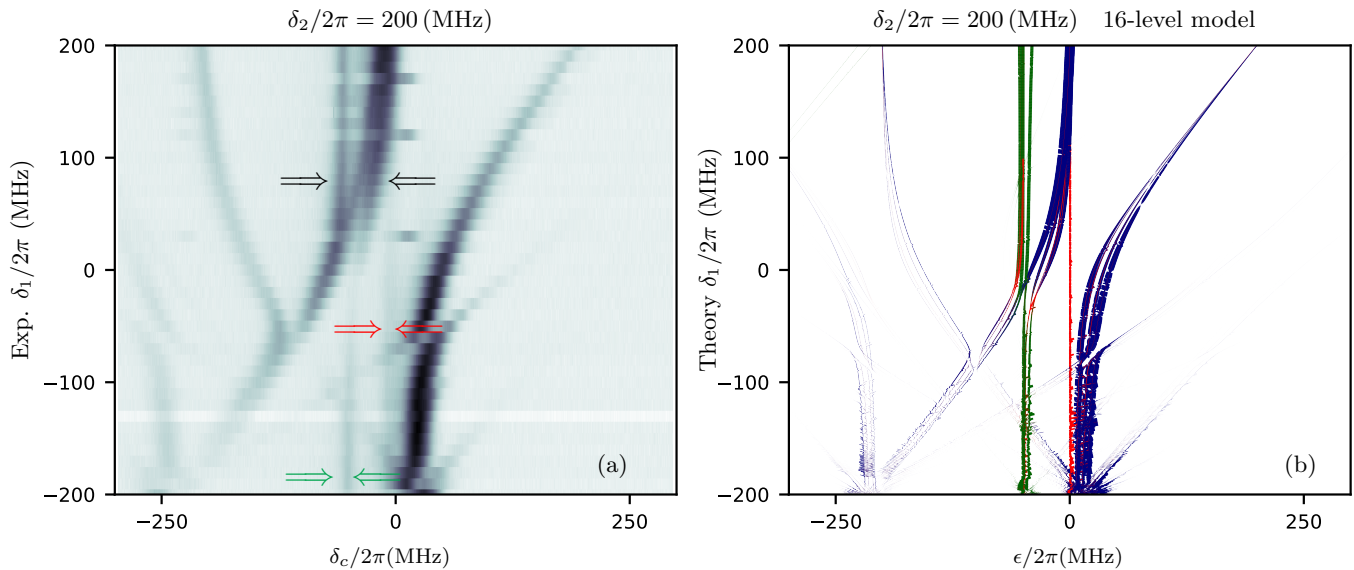

\subsection{16-level model: Magnetic sublevels of Rydberg fine structure}
The two-state and four-state models implicitly assume linear optical and RF polarizations. For imperfect RF polarization, faint additional lines appear in the experimental data. We confirm the locations of some of these additional features with a sixteen-level Hamiltonian that includes the magnetic subevels $m_J$ of the Rydberg fine-structure (see Appx.~\ref{sec:numbers}). In Fig.~\ref{figure:6}\,(b) we show results from a sixteen-state computation that includes the $m_J$ sublevels of the states $61D_{5/2}$, $61D_{3/2}$, $62P_{3/2}$, and $62P_{1/2}$. We highlight relevant features of our spectra that are reproduced in this extended model, below.

First, the green arrows in the figure highlight the $61D_{3/2}$ fine-state in the EIT spectra. 

Second, the red arrows indicate the contamination of the Floquet quasienergy levels with $|m_J|>1/2$ sublevels of the Rydberg states, that are not accounted for in the two-state and 4-state models. 
    
Third, the complex spectra in the vicinity of the black arrows are also polarization effects that are only reproduced in the 16-state model. The mixing between the $61D_{5/2}$ and $61D_{3/2}$ states appears to take place via different $m_J$ sublevels, as indicated by the overlap of colours in the numerical Floquet spectra in the vicinity of these features.

We note that the inclusion of the fine states and the $m_J$ states are necessary for explaining the slight asymmetry of the spectra in comparing, for instance, Fig.~\ref{figure:4}\,(a1) and Fig.~\ref{figure:4}\,(g1). As an additional point of interest, we note that the magnetic sublevels are not usually resolved in the standard Autler--Townes EIT spectra that we use for electrometry, and their behaviour in different experimental configurations contributes to our understanding of the Rydberg atomic system. A careful analysis of the relative strengths of the features that correspond to the different $m_J$ sublevels could be used to analyse the polarization of RF fields in the spirit of previously investigated vector electrometry schemes \cite{PhysRevLett.111.063001}.

\subsection{Details of theoretical modeling}\label{sec:numbers}
Here we give further parameters used in computing the four-level and sixteen-level spectra, computed using the ARC package \cite{sibalic_arc_2017}.

We show the numerical values used in calculation, both energy offsets in Tabs.~\ref{tab:freqs}, and $m_J$-resolved dipole strengths in Tab.~\ref{tab:dips}, for each available polarization. 

Our Rydberg EIT spectra reflect the overlap of the Floquet modes with the bare Rydberg state $61D_{5/2}$, and the state $61D_{3/2}$, which are both detected in our EIT scans.

\begin{table}[]
\begin{tabular}{|cc|}
\hline
\multicolumn{2}{|c|}{Atomic Energy Gaps ($/h$) } \\ \hline
\multicolumn{1}{|c|}{RF transition $61~D_{J5/2} - 62~P_{3/2} $} & $9.226$~GHz \\ \hline
\multicolumn{1}{|c|}{Fine Gap $61~D~({J=5/2} \leftrightarrow {J=3/2} )$} & $50.339$~MHz \\ \hline
\multicolumn{1}{|c|}{Fine Gap $62~P~({J=3/2} \leftrightarrow {J=1/2})$ }  & $415.72$~MHz  \\ \hline
\end{tabular}
\caption{\label{tab:freqs} 
Relevant atomic energy gaps, expressed in frequency.}
\end{table}

\begin{table*}[]
\begin{tabular}{|c|c|c|c|}
\hline
\hspace{0.5cm}Fine states \hspace{0.5cm}~& \hspace{0.5cm}$\sigma^-$ transitions ($\Delta m_J=-1$)\hspace{0.5cm}~ & \hspace{0.5cm}$\pi$ transitions ($\Delta m_J=0$)\hspace{0.5cm}~& \hspace{0.5cm}$\sigma^+$ transitions ($\Delta m_J=+1$)\hspace{0.5cm}~ \\ \hline
& $m_J = -3/2\rightarrow-5/2$ &$m_J = -3/2\rightarrow-3/2$ &$m_J = -3/2\rightarrow-1/2$ \\ \hline
$\wp_{62P_{3/2},61D_{5/2}}$ & $-3055$  & $1932$ & $-966$ \\ \hline
$\wp_{62P_{3/2},61D_{3/2}}$ & $0$  & $967$ & $-789$ \\ \hline
$\wp_{62P_{1/2},61D_{3/2}}$ & $0$ & $0$  & $0$ \\ \hline
&$m_J = -1/2\rightarrow-3/2$ &$m_J = -1/2\rightarrow-1/2$ &$m_J = -1/2\rightarrow1/2$ \\ \hline
$\wp_{62P_{3/2},61D_{5/2}}$ & $-2366$ & $2366$  & $-1673$ \\ \hline
$\wp_{62P_{3/2},61D_{3/2}}$ & $789$  & $322$  & $-911$ \\ \hline
$\wp_{62P_{1/2},61D_{3/2}}$ & $-2765$  & $2258$  & $-1596$ \\ \hline
&$m_J = 1/2\rightarrow-1/2$ &$m_J = 1/2\rightarrow1/2$ &$m_J = 1/2\rightarrow3/2$ \\ \hline
$\wp_{62P_{3/2},61D_{5/2}}$ & $-1673$  & $2366$ & $-2366$ \\ \hline
$\wp_{62P_{3/2},61D_{3/2}}$ & $911$  & $-322$ & $-789$ \\ \hline
$\wp_{62P_{1/2},61D_{3/2}}$ & $-1596$  & $2258$  & $-2765$ \\ \hline
&$m_J = 3/2\rightarrow1/2$ &$m_J = 3/2\rightarrow3/2$ &$m_J = 3/2\rightarrow5/2$ \\ \hline
$\wp_{62P_{3/2},61D_{5/2}}$ & $-966$  & $1932$ & $-3055$ \\ \hline
$\wp_{62P_{3/2},61D_{3/2}}$ & $789$  & $-967$ & $0$ \\ \hline
$\wp_{62P_{1/2},61D_{3/2}}$ & $0$  & $0$ & $0$ \\ \hline
\end{tabular}
\caption{\label{tab:dips} 
Relevant $\sigma^-$, $\pi$, and $\sigma^+$-transition dipole moments used in calculations, in atomic units of $ea_0$.}
\end{table*}

\bibliography{name}

\end{document}